\documentclass[prb,preprint,showpacs,showkeys,preprintnumbers,amsmath,amssymb]{revtex4-1}

\usepackage{amsmath,amssymb,bm,epsfig}
\usepackage{dcolumn}
\usepackage{bm}
\usepackage{feynmf}

\begin{document}


\title{Radiative Corrections to Quantum Sticking on Graphene}

\author{Sanghita Sengupta and Dennis P. Clougherty}

\affiliation{
Department of Physics\\
University of Vermont\\
Burlington, VT 05405-0125}

\date{\today}

\begin{abstract}
We study the sticking rate of atomic hydrogen to suspended graphene using four different methods that include contributions from processes with multiphonon emission.  We compare the numerical results of the atom self-energy obtained by: (1) the loop expansion of the atom self-energy, (2)  the non-crossing approximation (NCA), (3) the independent boson model approximation (IBMA), and (4) a leading-order soft-phonon resummation method (SPR).  The loop expansion reveals an infrared problem, analogous to the infamous infrared problem in QED.  The 2-loop contribution to the sticking rate gives a result that tends to diverge for large membranes.  The latter three methods remedy this infrared problem and give results that are finite in the limit of an infinite membrane.  We find that for micromembranes (sizes ranging 100 nm to 10 $\mu$m), the latter three methods give results that are in good agreement with each other and yield sticking rates that are mildly suppressed relative to the lowest-order golden rule rate. Lastly, we find that the SPR sticking rate decreases slowly to zero with increasing membrane size, while both the NCA and IBMA rates tend to a nonzero constant in this limit.   Thus, approximations to the sticking rate can be sensitive to the effects of soft-phonon emission for large membranes.

\end{abstract}

\maketitle
\section{Introduction}
There is renewed interest in the quantum mechanics of ultracold atoms near surfaces; work is underway to develop new quantum technologies such as chip-scale atomic clocks, quantum computers, and high-precision nanoscale sensors.  Understanding the dynamics of ultracold atoms near surfaces may give insights into optimizing the performance of these quantum devices.

There are additional practical considerations that motivate the study of ultracold atoms near surfaces; namely,  the ultracold regime offers some advantages for controlling chemical processes. Reaction pathways might be selected by controlling the quantum state of the reactants.  Thus,  a quantum theory of ultracold adsorption would be desirable for understanding heterogeneous catalysis at ultralow temperatures.

Although theoretical studies of ultracold surface adsorption began nearly eighty years ago, recent results on the adsorption or ``sticking'' of cold atomic hydrogen on a newly discovered 2-dimensional material, suspended graphene, have been controversial \cite{dpc12,LJcutoff}.  A numerical study of inelastic scattering from suspended graphene \cite{graphene} concluded that compared to sticking on graphite, sticking should be enhanced on 2-dimensional graphene.  The authors argued that sticking proceeds by the creation a single graphene phonon.  In contrast to 3-dimensional graphite, the vibrations of suspended graphene are essentially completely polarized normal to its surface.  Thus, the atom-phonon interaction is stronger for graphene than for graphite.

Another theoretical study found a very different result.  The author argued \cite{dpc13} that the frequency dependence of the atom-phonon interaction is also different for graphene than for graphite. An enhanced atom-phonon interaction at low frequencies resulting from the low dimensionality leads to a reduction of the sticking via a phonon orthogonality catastrophe.  An atom bound to the surface is accompanied by a deformation that involves a large number of low energy phonons.  Such a state has an exponentially small overlap with  graphene's vibrational ground state when the atom is in the gas phase. As a result, the sticking matrix element for a 1-phonon process is exponentially reduced and tends to vanish as the size of the graphene target becomes large.

In this work, we consider the effects of multiphonon sticking processes using diagrammatic perturbation theory.  We show that while the sticking rate in lowest order in the atom-phonon coupling is finite, the next term in the perturbation expansion tends to diverge as the size of the graphene target becomes large. This infrared divergence is rooted in the low frequency dependence of the atom-phonon coupling for flexural phonons.  Thus, the 1-phonon result can be a poor approximation to the total sticking rate.  This compels us to consider multiphonon contributions to the sticking rate.

To investigate the effects of the multiphonon emission on the adsorption rates, we consider three non-perturbative methods: (1) the independent boson model approximation (IBMA),  (2) the non-crossing approximation (NCA),  and (3) a leading-order soft phonon resummation method (SPR).  In the IBMA \cite{DPCPRB2016}, we replace in the diagrammatic expansion of the atom self-energy the Green function for the bound atom by the exact IBM Green function.  This approximation, equivalent to an infinite sum of diagrams containing the bound state atom-phonon vertices, includes multiphonon contributions to the atom self-energy.


In our second method, we use the non-crossing approximation (NCA) to calculate the atom self-energy.  This is equivalent to an infinite sum of ``rainbow" diagrams and also includes multiphonon contributions to the self-energy.  We iterate numerically the NCA nonlinear integral equation for the atom self-energy until self-consistency is achieved.    We find that both approximations yield finite adsorption rates in the limit of large graphene targets, and these adsorption rates are reduced relative to the first-order golden rule estimate.

In the third method, we include the interaction of the phonons with the bound atom in the unperturbed Hamiltonian, while the remaining atom-phonon interaction is treated perturbatively.  We sum over all emitted phonon states the dominant contributions to the sticking rate in the large membrane regime \cite{dpc16}.  We obtain a result that is in good agreement with NCA and IBMA results for micromembranes; however, for SPR, the sticking rate tends to zero as the size of the membrane becomes large.  The NCA and IBMA rates tend to a nonzero constant in this large membrane limit.

We have organized this paper in the following way: in Sec.~\ref{sec:PE}, we introduce our model Hamiltonian and describe the self-energy expansion resulting from Feynman-Dyson perturbation theory. The atom self-energy  captures the effects of interaction with the flexural phonons. We perform a systematic series expansion of the self-energy and show that to the lowest order in coupling, the transition rate is equal to that obtained by Fermi's golden rule; however, the higher order terms in the perturbative expansion suffer from divergences in the infrared limit leading us to the conclusion that approximations obtained by truncating the perturbation expansion are invalid for suitably large membranes in this model.

Our nonperturbative methods are introduced in Sec.~\ref{sec:RT}. These methods are tantamount to summations over particular classes of diagrams. These schemes take into consideration the strong coupling to low-energy phonons and eliminate the infrared singularities.   In Sec.~\ref{sec:results}, we compare the results of these approximations with the SPR result \cite{dpc16} that utilizes a coherent basis for the final phonon states.  We then discuss the effect of soft-phonon emission in cold atom adsorption to graphene and other 2D materials.

\section{Perturbative Expansion of the Self-energy}
\label{sec:PE}
We introduce the model Hamiltonian and discuss the relation between the atom self-energy and the sticking rate.  We obtain each term in the perturbation series utilizing Feynman rules for the model, and we examine the conditions for convergence of the series.

We begin with the Hamiltonian of the model that represents the interaction of a cold atom moving at normal incidence with a 2D elastic membrane. The previously derived \cite{dpc13} Hamiltonian is written as $H = H_{a}+H_{ph}+H_{bi}+H_{ki}$ where 
\begin{equation}\label{ham}
H_{a} = E_{k}c_{k}^{\dagger}c_{k}-E_{b}b^{\dagger}b 
\end{equation}
\begin{equation}\label{2}
H_{ph} = \sum_{q}\omega_{q}a_{q}^{\dagger}a_{q}
\end{equation}
\begin{equation}\label{3}
H_{ki} = -g_{kb}(c_{k}^{\dagger}b + b^{\dagger}c_{k})\sum_{q}(a_{q}+a_{q}^{\dagger})
\end{equation}
\begin{equation}\label{4}
H_{bi}=- g_{bb}b^{\dagger}b\sum_{q}(a_{q}+a_{q}^{\dagger})
\end{equation}

Here, $c_{k}$ $(c_{k}^{\dagger})$ annihilates (creates) an atom in the entrance channel $|k\rangle$ with energy $E_{k}$;  $b$ $(b^{\dagger})$ annihilates (creates) an atom in the bound state $|b\rangle$ with energy -$E_b$ in the static potential; $a_q$   $(a_q^{\dagger})$ annihilates (creates)  a flexural phonon with energy $\omega_q$;  $g_{kb}$ is the strength of a phonon-assisted atom transition between the continuum $|k\rangle$ and bound state $|b\rangle$; $g_{bb}$ is the coupling strength of the bound atom to the phonons \cite{dpc13,DPCPRB2016,dpc16}.

We now consider the atom-phonon interactions as perturbations with $H_{i} = H_{bi}+H_{ki}$.
For the  Hamiltonian in Eq.~\ref{ham}, the unperturbed Green functions for the atom in the continuum $|k\rangle$ and the bound state $|b\rangle$ are given as
\begin{equation}\label{GFK}
G^{(0)}_{kk}(E) = \frac{1}{E-E_{k} +i\eta} \ ,
\end{equation}
\begin{equation}\label{GFB}
G^{(0)}_{bb}(E) = \frac{1}{E+E_{b} + i\eta} \ ,
\end{equation}
while the phonon propagator is
\begin{equation}\label{9}
D(q,\omega)= \frac{2\omega_{q}}{\omega^{2}-\omega_{q}^{2} +i\eta} \ ,\eta \rightarrow 0^{+}.
\end{equation}

The atom self-energies $\Sigma_{kk}(E)$ and $\Sigma_{bb}(E)$ are obtained from application of the Feynman rules  for this model \cite{DPCPRB2016} (see Fig.~\ref{fig:FD1}). The sticking rate $\Gamma$ then follows \cite{DPCPRB2016} from the atom self-energy $\Sigma_{kk}$.  Thus,
\begin{equation}\label{TR}
\Gamma = -2Z(E_{k}) {\rm Im} \Sigma_{kk}(E_{k}) ,
\end{equation} 
where Z is the renormalization factor given by 
\begin{equation}
Z(E) = \bigg[1-\bigg({\partial {\rm Re}\Sigma_{kk}(E)\over \partial E}\bigg)\bigg]^{-1}
\end{equation}
In the next subsections, we obtain analytic expressions for the 1-loop and 2-loop self-energies.

 
\subsection{1-loop Self-Energy}
\label{sub:1se}

\begin{figure}[h!]
\includegraphics[width=0.95\columnwidth]{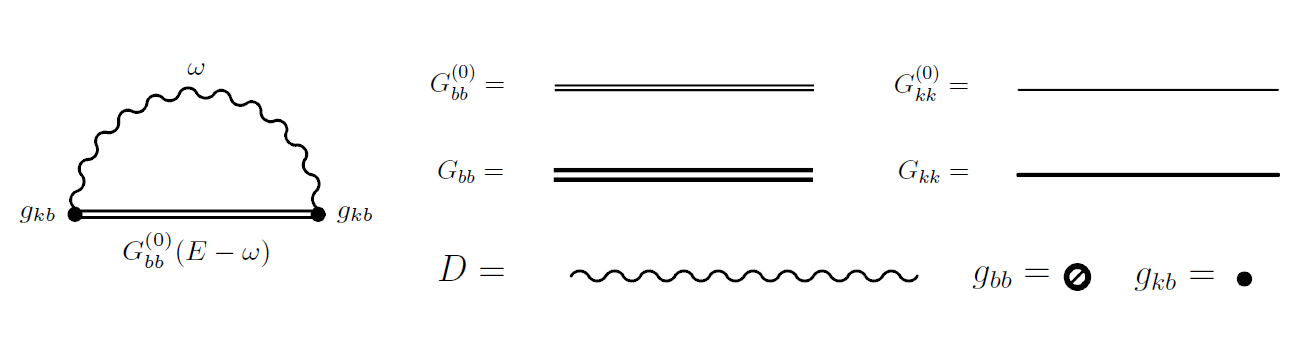} 
\caption{Feynman diagram for the 1-loop atom self-energy $\Sigma_{kk}^{(1)}$ (left). Symbolic elements used in the construction of diagrams are pictured (on right).}
\label{fig:FD1}
\end{figure}

We begin our calculations for the perturbation series with the derivation of the 1-loop atom self-energy $\Sigma_{kk}^{(1)}(E)$ (see Fig.~\ref{fig:FD1}). 

With our Feynman rules \cite{DPCPRB2016}, we find an analytical expression corresponding to Fig.~\ref{fig:FD1} 
\begin{equation}\label{Skk1}
\begin{split}
\Sigma_{kk}^{(1)}(E) &= ig_{kb}^{2}\int\limits_{-\infty}^{\infty}\frac{\mathrm{d\omega}}{2\pi}D(\omega)G_{bb}^{(0)}(E-\omega)\\
&\quad = \sum_{q}\frac{g_{kb}^{2}}{E+E_{b}-\omega_{q} + i\eta} \ , \eta\rightarrow 0^{+}
\end{split}
\end{equation}
(We work in natural units where $\hbar =$1.)

From the Wigner-Eckart theorem, we conclude that we need only consider axisymmetric vibrational modes for transitions to bound states that are also axially symmetric \cite{dpc13,DPCPRB2016}.  We evaluate the sum in Eq.~\ref{Skk1} in the continuum limit and obtain 
\begin{equation}\label{fullSkk1}
\begin{split}
\Sigma_{kk}^{(1)}(E) &= -g_{kb}^2\rho_{0}\ln\bigg|\frac{\omega_{D}-E-E_{b}}{E+E_b}\bigg|\\
&\quad -i\pi g_{kb}^2\rho_{0}\bigg[\theta(\omega_{D}-E-E_{b})-\theta(-E-E_{b})\bigg] \  , 
\end{split}
\end{equation}
where $\rho_{0}$ is the (constant) density of axisymmetric vibrational states and $\omega_{D}$ is the Debye frequency for the membrane.  $\Sigma_{kk}$ has an imaginary part  for $-E_{b}<E<\omega_{D}-E_{b}$ corresponding to atom transitions out of the continuum state to the bound state with the emission of a single phonon. Following Eq.~\ref{TR}, we obtain the first-order adsorption rate: 
\begin{equation}\label{RSkk1}
    \Gamma_{0} \approx 2\pi g_{kb}^{2}\rho_{0} \
\end{equation}
 for atom energies $E_k$ such that $E_k+E_b\le\omega_D$.
Also, we take $Z \approx 1$ which is valid for $g_{kb}^2\rho_0 \ll E_b$.

We conclude on the basis of Eq.~\ref{RSkk1} that at the 1-loop level, the transition rate is finite, proportional to $g_{kb}^{2}$ and is independent of the bound atom-phonon coupling $g_{bb}$. 

The sticking rate can also be calculated using Fermi's Golden Rule, with lowest order transition rate $\Gamma_{0}$ given by 
\begin{equation}\label{GR}
\Gamma_{0} = 2\pi \sum_{f}|\langle f|H_{i}|i\rangle|^{2}\delta(E_{f}-E_{i}) 
\end{equation}\\
with the initial and final states given as $|i\rangle = |k\rangle|0_{q}\rangle$ and $|f\rangle = |b\rangle|1_{q}\rangle$, respectively. The energies corresponding to the final and the initial state are $E_{f} = -E_{b}+\omega_{q}$ and $E_{i} = E_{k}$.  We then calculate the transition matrix element as
$\langle f|H_{i}|i\rangle =  -g_{kb}$.

Thus  sticking rate $\Gamma_0$ is
\begin{equation}\label{17}
\Gamma_{0} = 2\pi\sum_{q}g_{kb}^2\delta(-E_{b}+\omega_{q}-E_{k}) \ ,
\end{equation}
which, in the quasicontinuum approximation, becomes 
\begin{equation}\label{RGR}
\Gamma_{0} = 2\pi g_{kb}^2\rho_{0} \ .
\end{equation}
Therefore, we find that the 1-loop result is equal to transition rate given by Fermi's Golden rule (Eq.~\ref{RGR}) for $g_{kb}^2\rho_0 \ll E_b$.  We now examine the next term in the loop expansion of the self-energy. 

\subsection{2-loop Self-Energy}
\label{sub:2se}

The Feynman diagrams for the 2-loop terms of the atom self-energy $\Sigma_{kk}(E)$ are given in Fig. \ref{fig:2loop}. 
\begin{figure}[h!]
\includegraphics[width=.8\columnwidth]{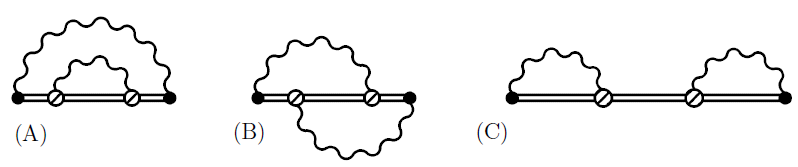} 
\caption{\label{fig:2loop} Feynman diagrams for the 2-loop terms for $\Sigma_{kk}$: (a) nested (or rainbow), (b) overlap, and (c) loop-after-loop.  }
\label{fig:2loop}
\end{figure}

The nested diagram is found to contain a power law divergence from soft phonons; however, for a finite-sized membrane, the vibrational  spectrum is cutoff at low frequencies.  We introduce $\epsilon$, the lowest vibrational frequency of the membrane, that scales as the inverse size of the membrane.  It serves as  a natural regulator for the infrared divergences.   We now calculate the $\epsilon$ dependence of the Feynman amplitudes in Fig.~\ref{fig:2loop}.  This cutoff dependence is summarized in Table I.

The nested diagram contributes $\Sigma_{kk}^{(2a)}$ to the atom self-energy
\begin{equation}\label{20}
\begin{split}
\Sigma_{kk}^{(2a)} &= i^2\int\int\frac{\mathrm{d}\omega}{2\pi}\frac{\mathrm{d}\omega'}{2\pi}\sum_{q}\sum_{q'}g_{kb}^{2}g_{bb}{^2}G^{(0)}_{bb}(E-\omega)\\
&\quad \times G_{bb}^{(0)}(E-\omega-\omega')G_{bb}^{(0)}(E-\omega)D(q,\omega)D(q',\omega')\\
&\quad = \sum_{q}\sum_{q'}\frac{g_{kb}^{2}g_{bb}^{2}}{(E+E_{b}-\omega_{q}+i\eta)^{2}}\\
&\quad \times \frac{1}{(E+E_{b}-\omega_{q}-\omega_{q'}+i\eta)} \ .
\end{split}
\end{equation}
In the quasicontinuum approximation, we evaluate the double integral and find that the real part diverges as a result of the double pole at $\omega=E+E_{b}$ and the imaginary part suffers from a linear divergence as $\omega\rightarrow 0$.

Similarly, one can derive analytical expressions for the overlap diagram
\begin{equation}\label{21}
\begin{split}
\Sigma_{kk}^{(2b)} &= i^2\int\int\frac{\mathrm{d}\omega}{2\pi}\frac{\mathrm{d}\omega'}{2\pi}\sum_{q}\sum_{q'}g_{kb}^{2}g_{bb}{^2}G_{bb}^{(0)}(E-\omega)\\
&\quad \times G_{bb}^{(0)}(E-\omega-\omega')G_{bb}^{(0)}(E-\omega')D(q,\omega)D(q',\omega')\\
&\quad = \sum_{q}\sum_{q'}\frac{g_{kb}^{2}g_{bb}^{2}}{(E+E_{b}-\omega_{q}+i\eta)}\\
&\quad \times \frac{1}{(E+E_{b}-\omega_{q}-\omega_{q'}+i\eta)(E+E_{b}-\omega_{q'}+i\eta)} \ .
\end{split}
\end{equation}
In the quasicontinuum approximation,  this integral contains a logarithmic divergence for both the real and imaginary parts  from low-frequency phonons ($\omega\rightarrow0$). In Table I, we specify the $\epsilon$ dependence from the most divergent terms contained in the three 2-loop diagrams.


\begin{table}[h!]\label{table}
\begin{center}
    \renewcommand{\arraystretch}{2}
  \begin{tabular}{  c  c  c     }
   \hline\hline 
    Feynman Diagram & Re $\Sigma_{kk}^{(2)}$ & Im $\Sigma_{kk}^{(2)}$ \\ 
    \hline
    Nested  & $-\infty$  &  -$\frac{\pi g^{2}}{\epsilon}$ \\ 
    Overlap & $\frac{g^{2}}{E_b}\ln\frac {\omega_{D}}{E_b}\ln\frac{E_b}{\epsilon}$ & $-\frac{2\pi g^{2}}{E_{b}}\ln\frac{\omega_D}{\epsilon}$ \\
    Loop-after-loop & $\frac{g^{2}}{E_{b}}\bigg[\bigg(\ln \frac{\omega_{D}}{E_{b}}\bigg)^{2}-\pi^{2}\bigg] $ & $\frac{2\pi g^{2}}{E_b}\ln\frac{\omega_{D}}{E_b}$\\
    \hline \hline
  \end{tabular}
\end{center}
\caption{\label{tab:div} The $\epsilon$ dependence of the 2-loop contributions to the atom self-energy $\Sigma^{(2)}_{kk}(E)$. We define $g = g_{kb}g_{bb}\rho_{0}$. The infrared frequency cutoff $\epsilon$   is inversely proportional to the membrane size $\epsilon \sim v_{s}/L$, where $v_{s}$ is the transverse speed of sound and $L$ is the size of the membrane. The nested diagram has the most divergent contribution to the atom self-energy. }
\end{table}


While the loop-after-loop diagram is finite,  we find that the 2-loop nested and overlap diagrams contain contributions that tend to diverge with increasing membrane size.  Hence, the loop expansion becomes unreliable for suitably large membranes. This is an phonon version of the infrared problem in QED \cite{dpc16}.   The breakdown of perturbation theory leads us to consider alternative methods.  We discuss the results of two resummation methods that remedy this infrared problem for large membranes. 

\section{Resummation Methods}
\label{sec:RT}

In our previous work on the studies of adsorption of cold atoms on high-temperature 2D elastic membranes  \cite{DPCPRB2016},  we attempted to go beyond the truncated loop expansion by considering a partial resummation technique in the spirit of the well-known exact solution of the independent boson model (IBM).  Henceforth, we refer to this replacement of $G^{(0)}_{bb}$ in the diagrammatic expansion with the exact IBM Green function $G_{bb}^{\rm IBM}$ as the independent boson model approximation (IBMA).  This replacement has similarities to the Bloch-Nordsieck scheme \cite{Bloch:1937pw} used in hot QED and QCD plasmas\cite{qed2,qed3}.

In this work, in addition to the IBMA, we employ another resummation technique that corresponds to the summation of all rainbow diagrams- the non-crossing approximation (NCA). The reason behind using NCA is twofold: firstly, for the zero temperature case, we see that the the leading order divergence is rooted in the rainbow (nested) diagram. Hence, we aim to sum all the rainbow diagrams to infinite order with the motive of curing the leading-order divergent contributions.  Also, the NCA is a resummation technique which is well suited to account for strong divergences. Secondly, there are  successful applications of the NCA in the other fields of many-body physics such as the Anderson impurity problem \cite{haule,otsuki}, quantum dot transport \cite{wingreen}, and role of phonon interactions in the Anderson-Holstein model, in a quantum antiferrromagnet and in Holstein-Hubbard model \cite{chen,kar,werner}.

In the next subsection, we introduce the IBMA and use it to calculate the atom self-energy $\Sigma_{kk}$ in the infrared limit $\epsilon\rightarrow 0$.


\subsection{Independent Boson Model Approximation}
\label{sub:IBMA}

The model in Eq.~\ref{ham} can be viewed as a generalization of the IBM with two coupling constants $g_{kb}$ and $g_{bb}$. Our primary interest is in describing ultracold atoms. Thus we focus on the  regime where $g_{bb} \gg g_{kb}$, as $g_{kb}$ is reduced by the effect of quantum reflection for ultracold neutral atoms \cite{dpc92}. We aim to derive an exact solution in the stronger channel ($|b\rangle$) compared to the continuum ($|k\rangle$) which we treat perturbatively \cite{DPCPRB2016}. An exact solution for the stronger channel with atom-phonon coupling $g_{bb}$ is obtained in terms of the IBM. We begin with the IBM Hamiltonian which is contained in our model Hamiltonian. The IBM only considers the interaction of the bound atom of energy $-E_{b}$ with the flexural phonons with energy $\omega_{q}$.  Thus, 
\begin{equation}\label{hamibm}
H_{IBM} = -E_{b}b^{\dagger}b + \sum_{q}\omega_{q}a_{q}^{\dagger}a_{q} - g_{bb}b^{\dagger}b\sum_{q}(a_{q}+a_{q}^{\dagger})
\end{equation}

The exact time-dependent Green function  of the IBM $G_{bb}^{\rm IBM}(t)$ is \cite{mahan}
\begin{equation}\label{ibmG}
G_{bb}^{\rm IBM}(t) = -i e^{-it(-E_{b}-\Delta)}e^{-\phi(t)}
\end{equation}
where $\phi(t) = \sum_{q}g_{bb}^{2}(1-e^{-i\omega_{q}t})/\omega_{q}^{2}$ and the polaron shift $\Delta$  is given as\cite{mahan},
\begin{equation}\label{Delta}
\Delta = \sum_{q}\frac{g_{bb}^{2}}{\omega_{q}} \to \int_{\epsilon}^{\omega_{D}}\frac{g_{bb}^{2}\rho_{0}}{\omega}\mathrm{d}\omega
\end{equation}

In the quasicontinuum approximation, we rewrite Eq.~\ref{ibmG} as
\begin{equation}\label{cibmG}
\begin{split}
G_{bb}^{\rm IBM}(t) &= -i\exp[it(E_{b}+\Delta)] \exp\bigg[-ig_{b}\int_{\epsilon}^{\omega_{D}}\bigg(\frac{\mathrm{d}\omega \sin\omega t}{\omega^{2}}\bigg)\bigg]\\
&\quad\times \exp\bigg[-g_{b}\int_{\epsilon}^{\omega_{D}}\mathrm{d}\omega\bigg(\frac{1-\cos\omega t}{\omega^{2}}\bigg)\bigg]
\end{split}
\end{equation}
where we have defined $g_{b}\equiv g_{bb}^{2}\rho_{0} $.
The Fourier transform of $G_{bb}(t)$ is 
\begin{equation}\label{FIBM}
\begin{split}
G_{bb}^{\rm IBM}(E_s) &= -i\int_{0}^{\infty}\bigg[e^{it(E_s+\Delta)}\\
&\quad \times \exp\bigg(-ig_{b}\int_{\epsilon}^{\omega_{D}}\frac{\mathrm{d}\omega \sin\omega t}{\omega^{2}}\bigg) \\
&\quad \times \exp\bigg[-g_{b}\int_{\epsilon}^{\omega_{D}}\mathrm{d}\omega\bigg(\frac{1-\cos\omega t}{\omega^{2}}\bigg)\bigg]\bigg]\mathrm{d}t
\end{split}
\end{equation}
where the energy $E_s$ is defined as $E_{s} = E + E_{b}$.

\begin{figure}[h!]
\includegraphics[width=.8\columnwidth]{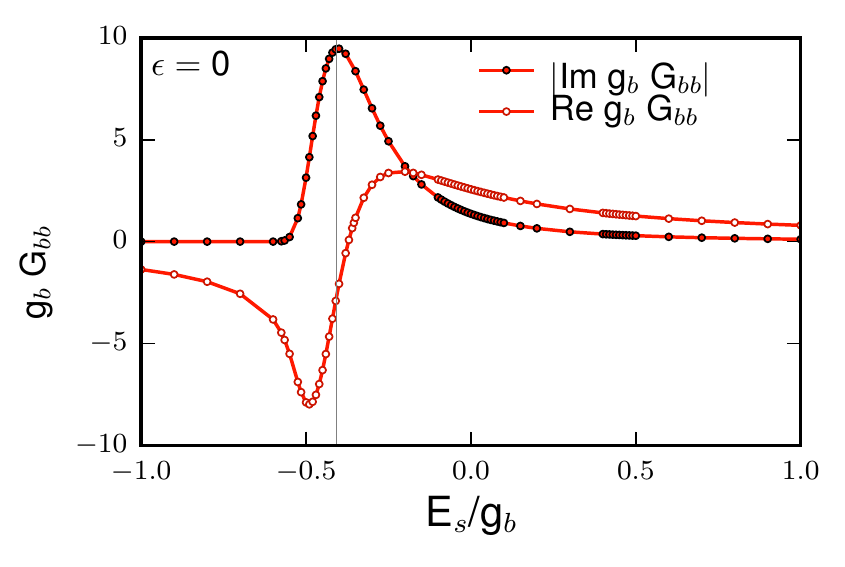} 
\caption{\label{fig:PIBM} The real and imaginary parts of scaled IBM Green function $g_b G_{bb}^{\rm IBM}(E_s/g_b)$ vs $E_s/g_b$ for $\epsilon=0$.  $G_{bb}^{\rm IBM}(E_s/g_b)$ is a smooth, complex-valued function in the limit $\epsilon\to 0$.}
\end{figure}

The above integral is computed numerically and gives convergent, finite results for both the real and imaginary parts of $G_{bb}(E_{s})$ in the infrared limit $\epsilon \rightarrow 0$.  This is somewhat surprising at first blush, as the polaron shift in Eq.~\ref{Delta} clearly diverges logarithmically as $\epsilon\to 0$.  There is however a cancellation in the phases that gives a finite limit.  It is convenient to rewrite Eq.~\ref{FIBM} in the scaling form $g_b G_{bb}(E_s/g_b)$.  In the limits of $\omega_D\to\infty$ and $\epsilon\to 0$, this scaling form is a universal function.  Fig.~\ref{fig:PIBM} shows the imaginary and the real parts of the scaled Green function $g_b G_{bb}(E_s/g_b)$ as function of $E_s/g_b$.  We note that the imaginary part, proportional to the density of states for the bound atom coupled to the flexural phonons, has a broad, asymmetric peak centered at $E_s/g_b\approx -1+\gamma$ where $\gamma$ is the Euler-Mascheroni constant.  No weight of the quasiparticle (delta function) density of states in ${\rm Im} G_{bb}(E_s)$ remains when $\epsilon=0$.  The real part of $G_{bb}(E_s/g_b)$ vanishes approximately at the energy corresponding to the peak in the imaginary part.

Under the IBMA, we calculate the atom self-energy $\Sigma_{kk}$ by replacement of the bare bound state Green function $G_{bb}(E_s)$ in the diagram of Fig.~\ref{fig:FD1} by the IBM Green function $G_{bb}^{\rm IBM}(E_s)$. Owing to the relative strength of atom-phonon couplings, the infrared behavior produced by higher order processes in $g_{kb}$ will be small and thus can be neglected \cite{DPCPRB2016}. Hence, we can restrict our calculations for the self-energy $\Sigma_{kk}^{\rm IBM}$ at the first loop which would contain all orders in the strong coupling $g_{bb}^{2}$ but only to the lowest order in $g_{kb}^{2}$. 

The imaginary part of the atom self-energy is consequently evaluated by numerically integrating the following expression
\begin{equation}\label{sIBM}
\mathrm{Im}  \Sigma_{kk}^{\mathrm{IBM}} (E) = -\rho_0 g_{kb}^{2}\int _{\epsilon}^{\omega_{D}}\mathrm{Im} G_{bb}^{\mathrm{IBM}}(E_{s}-\omega)\mathrm{d}\omega
\end{equation}
Because ${\rm Im} G_{bb}^{\rm IBM}(E_s)$ is a smooth, finite, well-behaved function, the integral in Eq.~\ref{sIBM} converges in the infrared limit $\epsilon\rightarrow 0$. 

The IBMA sticking rate $\Gamma^{\rm IBM}$ is obtained by combining Eqs.~\ref{TR} and \ref{sIBM}, yielding
\begin{equation}\label{IBMrate}
 \Gamma^{\mathrm{IBM}} (E) = 2\pi g_{kb}^{2}\rho_0\int _{\epsilon}^{\omega_{D}}\rho_B(E_{s}-\omega) d\omega
\end{equation}
where $\rho_B(E)\equiv -\frac{1}{\pi}{\rm Im} G_{bb}^{\rm IBM}(E)$ is the bound atom density of states for $H_{\rm IBM}$.  $\rho_B(E)$ is subject to a sum rule and must, when integrated over all $E$, give 1.  (We verified this sum rule numerically as a partial check of our results.)  By comparing Eqs.~\ref{GR} and \ref{IBMrate}, we conclude that $\Gamma^{\rm IBM}\le \Gamma_0$.  Our numerical results for the IBMA sticking rate are discussed in Sec.~\ref {sec:results}. In our next subsection, we turn to NCA to calculate the sticking rate.


\subsection{Non-Crossing Approximation}
\label{sub:NCA}
\begin{figure}[h]
\includegraphics[width=.5\columnwidth]{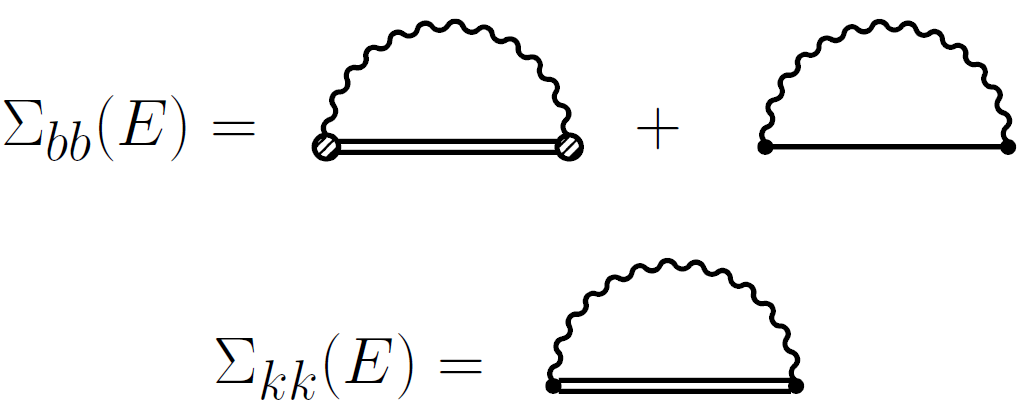} 
\caption{Diagrams in the NCA.   Evaluation of these diagrams results in a set of coupled nonlinear integral equations that may be solved self-consistently.  This is equivalent to a sum of all diagrams with no phonon lines crossing.} 
\label{fig:tFDNCA}
\end{figure}

The one-loop approximation to the atom self-energy $\Sigma_{kk}(E)$  (Fig.~\ref{fig:FD1}) depends on the bound atom Green function $G_{bb}^{(0)}$.  We can improve on this approximation by replacing the  bare bound atom Green function with the dressed bound atom Green function $G_{bb}$.  However, $G_{bb}$ depends on the bound atom self-energy $\Sigma_{bb}$ which is unknown.  That leads us to  consider the set of one-loop diagrams pictured in Fig.~\ref{fig:tFDNCA}.

Evaluation of the one-loop diagrams for the atom self-energies $\Sigma_{bb}$ and $\Sigma_{kk}$ gives the following coupled, nonlinear integral equations
\begin{equation*}
\begin{split}
\Sigma_{bb}(E) &= \sum_{q}\frac{g_{bb}^2}{E+E_{b}-\omega_{q}-\Sigma_{bb}(E-\omega_{q})}\\
&\quad + \sum_{q}\frac{g_{kb}^2}{E-E_{k}-\omega_{q}-\Sigma_{kk}(E-\omega_{q})} \ ,
\end{split}
\end{equation*}
\begin{equation}\label{NCA0c}
\Sigma_{kk}(E) = \sum_{q}\frac{g_{kb}^2}{E+E_{b}-\omega_{q}-\Sigma_{bb}(E-\omega_{q})} \ .
\end{equation}

\begin{figure}[h]
\includegraphics[width=.6\columnwidth]{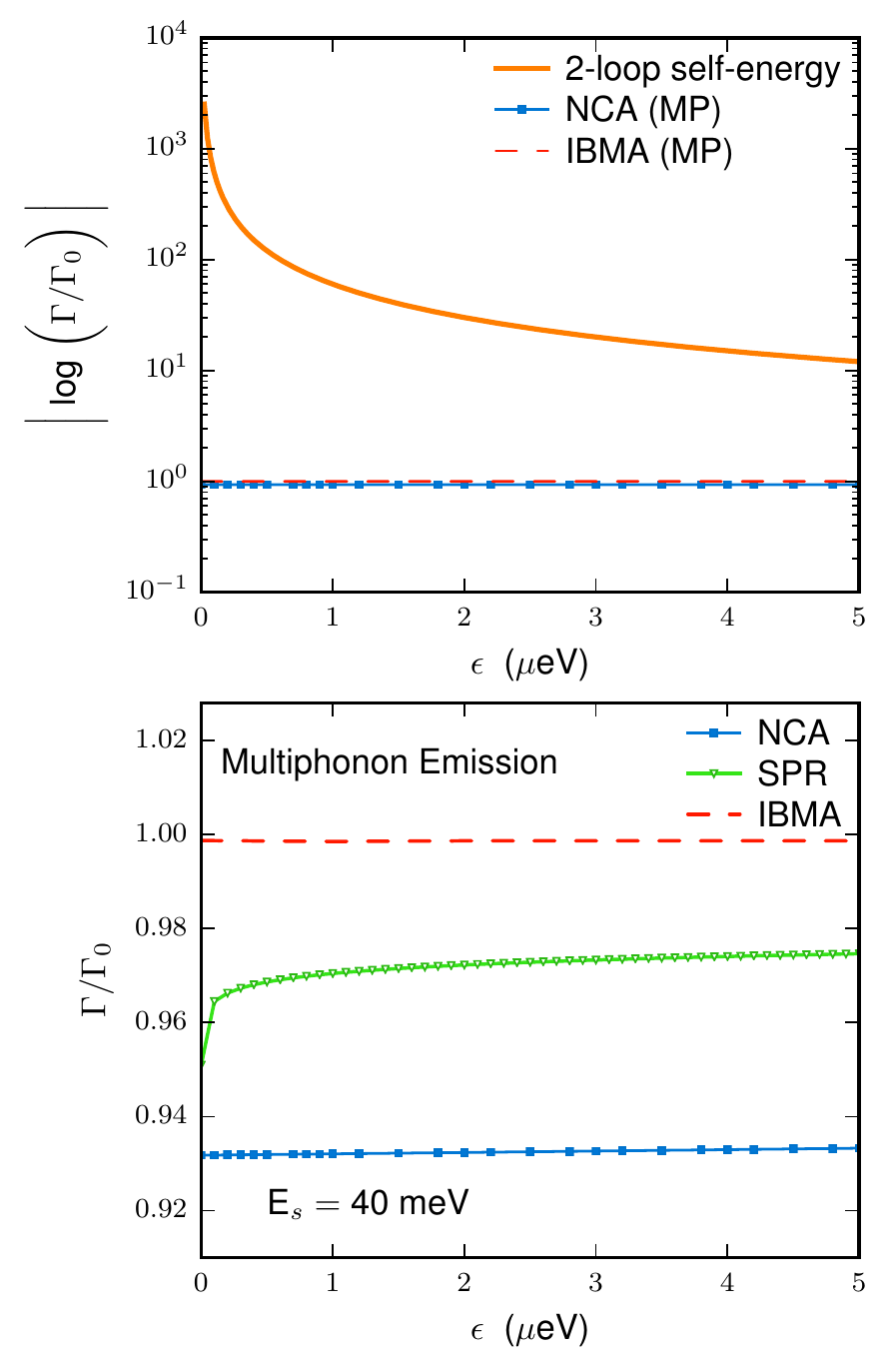} 
\caption{\label{fig:comp} Dependence of the (normalized) adsorption rate $\Gamma/\Gamma_0$ with IR cutoff $\epsilon$ (top panel). The two-loop approximation to $\Gamma$ diverges as $\epsilon\to 0$. Under NCA and IBMA, $\Gamma/\Gamma_0$ converges to a constant in this limit (bottom panel).  In SPR\cite{dpc16},  $\Gamma/\Gamma_0$ behaves asymptotically as $\ln^{-2}(\omega_D/\epsilon)$ as $\epsilon\to 0$.}
\end{figure}

\begin{figure}[h]
\includegraphics[width=.8\columnwidth]{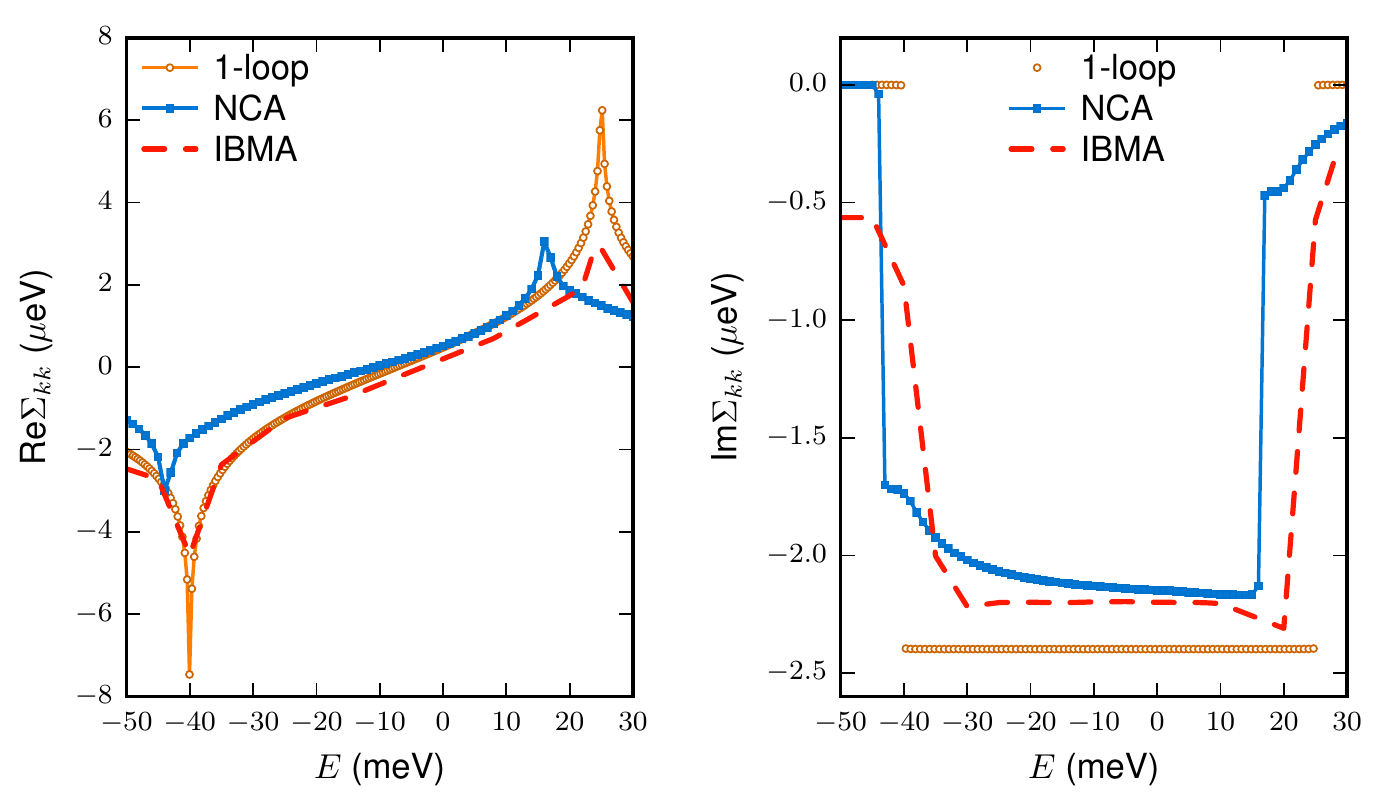} 
\caption{\label{nca-skk} Left panel: Real part of the atom self-energy ${\rm Re}\ \Sigma_{kk}(E)$ vs energy $E$.  Results are plotted for $\epsilon=0$, $g_{kb}^2\rho_0=0.7\ \mu$eV using three different approximation methods: one-loop self-energy $\Sigma_{kk}^{(1)}$, NCA, and IBMA. Right panel: Imaginary part of the atom self-energy ${\rm Im}\ \Sigma_{kk}(E)$ vs energy $E$. The sticking rate is obtained from self-energy using Eq.~\ref{TR}.}
\end{figure}

We numerically solve this set of equations by iteration until self-consistency has been achieved.  The numerical solutions to the atom self-energies were  found  to smoothly converge to a limiting form as $\epsilon\to 0$.   Fig.~\ref{nca-skk} shows the real and imaginary parts of the atom self-energy $\Sigma_{kk}$ as a function of $E$ using NCA. Using Eq.~\ref{TR}, we then obtain the adsorption rate under NCA.  The energy dependence of the calculated adsorption rate under the NCA is plotted in Fig.~\ref{fig:bound}.  The dependence of the NCA sticking rate with IR cutoff $\epsilon$ at fixed energy $E_s$ is plotted in Fig.~\ref{fig:comp}. 

\section{Adsorption of atomic hydrogen on graphene}
\label{sec:results}
We take the following for the numerical values of the parameters for the model of Eq.~\ref{ham}: $g_{bb}^2\rho_0=0.06$ meV,  $g_{kb}^2\rho_0=0.5-10\ \mu$eV, $\omega_D=65$ meV,  $E_b=40$ meV, and the transverse speed of sound in graphene is $v_{s} = 6.64 \times 10^{3}$ m/s.  We consider micromembranes whose size ranges from 100 nm to 10 $\mu$m.  
These values have been previously used to model \cite{dpc13,graphene,DPCPRB2016}  ultracold atomic hydrogen impinging at normal incidence on a suspended, circular sample of graphene.

We compare the numerical results of the sticking rate obtained by: (1) the loop expansion of the atom self-energy, (2)  the non-crossing approximation (NCA), (3) the independent boson model approximation (IBMA), and (4) a leading-order soft-phonon resummation method (SPR) \cite{dpc16}.

While the one-loop sticking rate gives a result that in the regime of interest is tantamount to the result from Fermi's golden rule, the two-loop sticking rate diverges linearly with increasing membrane size (or equivalently, with vanishing IR cutoff $\epsilon\to 0$).  This divergence is seen in the top panel of Fig.~\ref{fig:comp}.  We conclude that approximations based on truncation of the loop expansion break down for sufficiently large membranes.  

As anticipated, the IBMA sticking rate is slightly less than the golden rule rate $\Gamma_0$ over the range of IR cutoff $\epsilon$ considered (see lower panel of Fig.~\ref{fig:comp}). The reduction in sticking under IBMA was a result of the smearing of the bound atom density of states through the interaction with the phonons.   In contrast to the two-loop result, the IBMA rate remains finite in the limit of vanishing IR cutoff $\epsilon\to 0$.  We conclude that sticking in the IBMA which includes the spontaneous emission of an arbitrary number of phonons remedies the infrared problem in the loop expansion.  

The NCA sticking rate is also suppressed relative to $\Gamma_0$ and is also consistently lower than the IBMA sticking rate over the range of parameters used (see the lower panel of Fig.~\ref{fig:comp}).  The NCA further broadens the bound atom density of states through the inclusion of virtual transitions of the atom back to the continuum.  This further depresses the sticking rate in comparison to the IBMA.  We note that the NCA also gives a finite rate as $\epsilon\to 0$.

\begin{figure}[h]
\includegraphics[width=.55\columnwidth]{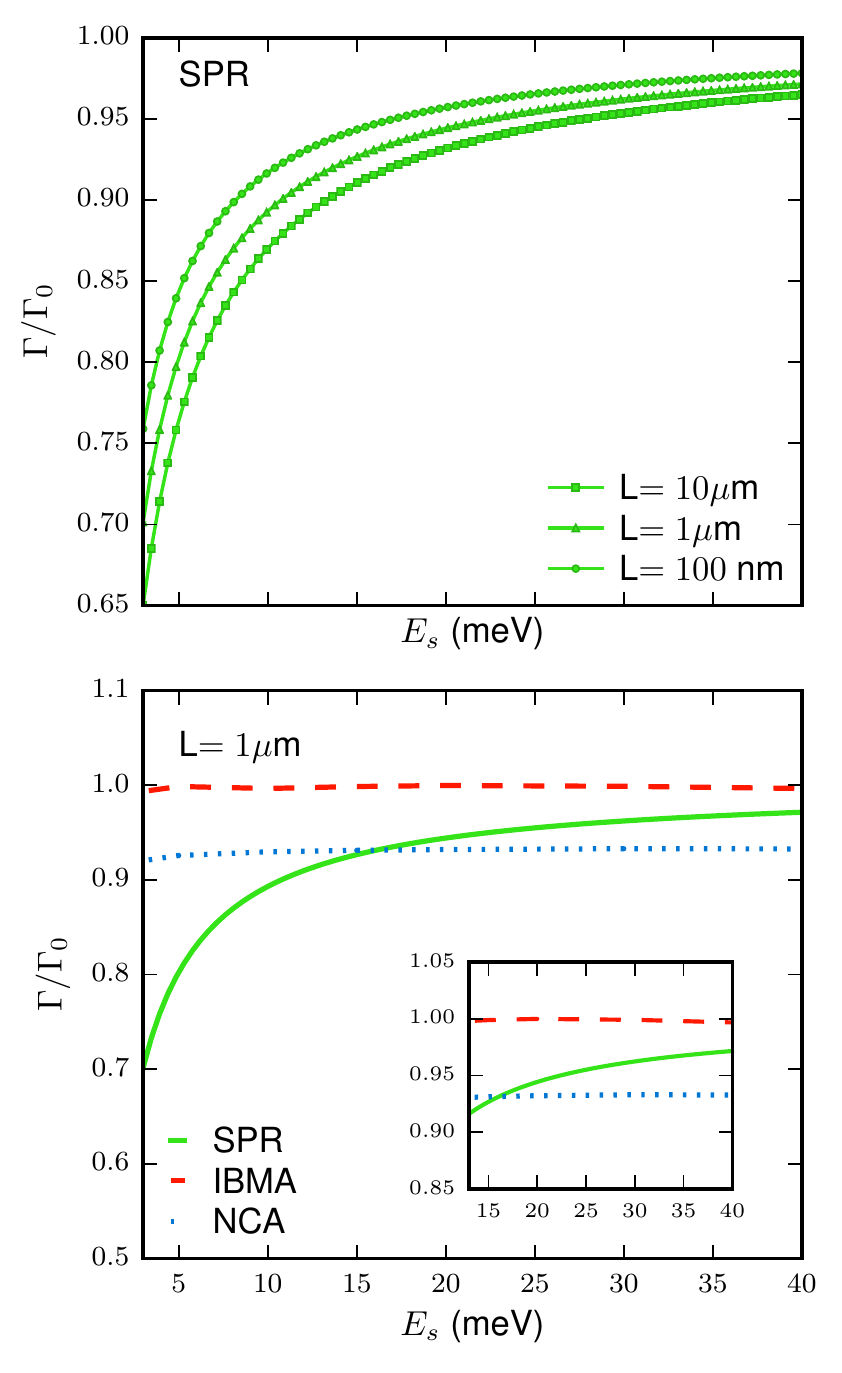} 
\caption{\label{fig:bound} Top panel: Normalized adsorption rate $\Gamma/\Gamma_0$ from the SPR method vs $E_s$ for various membrane sizes ranging 100 nm to 10 $\mu$m. Bottom panel: Normalized adsorption rate $\Gamma/\Gamma_0$ from SPR, IBMA and NCA vs $E_s$ for a 1 $\mu$m micromembrane.  For shallow bound states, the SPR result shows appreciable suppression in sticking relative to the other methods. However, for deep bound states, the three methods (IBMA, NCA and SPR)  give sticking rates that are in good agreement with each other. }
\end{figure}

Lastly, we plot the sticking rate under the SPR method \cite{dpc16} in Fig.~\ref{fig:comp} for comparative purposes.  We note that over the parameter range for micromembranes, the SPR sticking rate lies in between the IBMA and NCA rate.  Thus, there is good agreement with the three methods for micromembranes; however, the $\epsilon$ behavior of the SPR rate differs from both the NCA and IBMA rate.  The SPR method remedies the infrared problem, but in contrast to the IBMA and NCA rate, the SPR rate vanishes logarithmically \cite{dpc16} with $\Gamma\sim \Gamma_0 \ln^{-2}(\omega_D/\epsilon)$ as $\epsilon\to 0$.  The lower panel of Fig.~\ref{fig:bound} shows that for sufficiently low binding energy $E_b$, the SPR rate drops below the NCA rate.  As the SPR method is asymptotically exact as $\epsilon\to 0$, we anticipate that for weakly bound adsorbates the monotonic suppression of the sticking rate with increasing membrane size might be experimentally accessible.

In summary, we have studied the adsorption of ultracold atoms to a 2D elastic membrane. We have shown that the loop expansion produces a series that is divergent as $\epsilon\to 0$.  This infrared problem is similar to the infamous infrared problem of quantum electrodynamics where a scattered particle radiates an infinite number of soft photons. In QED, the infrared problem has been treated in several ways, including: (1) Bloch-Nordsieck theory \cite{Bloch:1937pw,Matiqed1} where the soft-photon corrections are arranged as an expansion in the photon frequency ($\hbar\omega/c\Delta p$ where $\Delta p$ is the momentum change from scattering) rather than the fine structure constant $\alpha$; (2) Kinoshita's cancellation of real and virtual corrections in perturbation theory \cite{kinoshita}; and (3) use of a photon coherent state basis \cite{IR-coherent}. Unencumbered by the constraints of Lorentz covariance in our model, the study of the phonon infrared problem may help to provide a more complete understanding of the issues in the QED counterpart.

In our work, both the IBMA and NCA replace the bare bound atom Green function with a propagator that includes atom-phonon interactions.  This dressed propagator leads to a broadened density of states for the bound atom and eliminates infrared divergences in the probability amplitudes.  The SPR method has similarity with the use of coherent states in QED \cite{IR-coherent}.  In the SPR method, the inclusion of the interaction of the bound atom with the phonons in the unperturbed Hamiltonian leads to the use of a coherent state basis for the phonons in the evaluation of the sticking amplitudes.  All three of these methods include the spontaneous emission of an arbitrary number of phonons.  

Our numerical results for micromembranes show that  the IBMA, NCA and SPR methods give results that are in good agreement with each other and yield sticking rates that are mildly suppressed relative to the lowest-order golden rule rate. All three methods produce results that are free of infrared divergences.  However, the SPR sticking rate decreases slowly to zero with increasing membrane size, while both the NCA and IBMA rates tend to a nonzero constant as $\epsilon\to 0$.   We conclude that  approximations to the sticking rate can be sensitive to the effects of soft-phonon emission for large membranes and that cold atom adsorption on a membrane might be viewed as a finite-size effect.


\bibliographystyle{apsrev4-1}
\bibliography{qs}

\end{document}